\documentclass[a4paper,12pt]{article}
\usepackage[utf8]{inputenc}

\usepackage{enumitem}
\usepackage{fullpage}
\usepackage{amsfonts}
\usepackage{amsmath}
\usepackage{mathtools}
\usepackage{tcolorbox}
\usepackage{graphicx}
\usepackage[bottom]{footmisc}
\usepackage{nccmath}
\usepackage{fancyhdr}
\usepackage{cite}
\usepackage{lmodern}
\usepackage{color}   %May be necessary if you want to color links
\usepackage{hyperref}
\usepackage[affil-it]{authblk}
\usepackage[titles]{tocloft}

\hypersetup{
    colorlinks,
    citecolor=blue,
    filecolor=black,
    linkcolor=blue,
    urlcolor=blue,
    linktocpage
}

\fancypagestyle{plain}{%
  \fancyhf{}% clears all header and footer fields
  \fancyfoot[C]{--~\thepage~--}%
}

\numberwithin{equation}{section}
\normalsize

\title{
\boldmath{Constraining the weights of Stokes Polytopes using BCFW recursions for $\phi^4$}}

\author{\normalsize
Ishan Srivastava \thanks{Email: ishan.alld@gmail.com}}
\affil{Chennai Mathematical Institute,\\ H1 SIPCOT IT Park, Siruseri, Chennai 603103, India}
\date{\vspace{-5ex}}
\begin{document}
\maketitle

\begin{abstract}
The relationship between certain geometric objects called polytopes and scattering amplitudes has revealed deep structures in QFTs. It has been developed in great depth at the tree- and loop-level amplitudes in $\mathcal{N}=4~\text{SYM}$ theory and has been extended to the scalar $\phi^3$ and $\phi^4$ theories at tree-level. In this paper, we use the generalized BCFW recursion relations for massless planar $\phi^4$ theory to constrain the weights of a class of geometric objects called Stokes polytopes, which manifest in the geometric formulation of $\phi^4$ amplitudes. 
We see that the weights of the Stokes polytopes are intricately tied to the boundary terms in $\phi^4$ theories.
We compute the weights of $N=1,2$, and $3$ dimensional Stokes polytopes corresponding to six-, eight- and ten-point amplitudes respectively. We generalize our results to higher-point amplitudes and show that the generalized BCFW recursions uniquely fix the weights for an $n$-point amplitude.
\end{abstract}
\vfill
\newpage

\tableofcontents

\rule{15cm}{0.1em}

\section{Introduction}
\label{sec:intro}
In recent years a lot of work has been done in understanding both the analytic and geometric structure of scattering amplitudes in various classes of quantum field theories \cite{1,2,3,4}. On-shell methods such as the BCFW recursion relations \cite{6,7,8,9,10} have been not only extremely successful in simplifying the calculations of an infinite class of amplitudes but also revealed deep connections between physics and broad areas of mathematics such as Algebraic Geometry and Combinatorics. In addition, generalizations of the BCFW recursion relations have been successfully devised and are used to incorporate the boundary contributions, which correspond to the pole at infinity, in a broad class of QFTs \cite{9,10}.

Furthermore, the seminal work of \cite{1} dubbed as the `Amplituhedron' program has been greatly successful in developing a geometric formulation of the $\mathcal{N}=4$ SYM amplitudes. An essential feature in this formulation was that the scattering amplitudes were considered not as functions of particle momentum but rather as differential forms on certain auxiliary spaces. It was shown that there is a relation between these differential forms and certain positive geometries, which completely encapsulates the $\mathcal{N}=4$ SYM amplitudes. A remarkable feature of this approach is that it is gauge-invariant and makes no reference to underlying principles of standard formulation of QFTs such as locality and unitarity, which are emergent in the geometric formulation.

In \cite{2,3} the authors extended this program to scalar $\phi^3$ and $\phi^4$ theories. In particular, for the massless planar tree-level $\phi^3$ amplitudes, it was shown in \cite{2} that there is a precise relation between so-called planar scattering forms on kinematic space and a polytope known as Associahedron. Further in \cite{3,4}, the authors developed a similar formulation for the massless planar $\phi^4$ amplitudes at tree-level and in \cite{5} it was extended to the generalized case of $\phi^p$ interactions. It was shown that the planar $\phi^4$ amplitudes can be obtained from the geometry of an object known as Stokes polytopes. However, it was found that the calculation of $\phi^4$ amplitudes from the geometry of Stokes polytopes presents a peculiarity. The peculiarity lies in the fact that for a given number of particles $n$, there does not exist a unique Stokes polytope which completely determines the $\phi^4$ amplitudes. In contrast, there exists a unique Associahedron for a given $n$, which completely encapsulate the tree-level $\phi^3$ amplitudes. Each Stokes polytope contains only partial information about the complete $\phi^4$ amplitudes. 

In order to determine the complete $\phi^4$ amplitudes, a weighted sum over all the Stokes polytopes is taken, and in general, these weights are not equal. This problem is made simpler by the fact that a cyclic permutation of the labels of only a few Stokes polytopes, referred to as the primitive Stokes polytopes, determine all the Stokes polytopes of a given dimension. As a result of this fact, the weights can be parametrized only by the primitive quadrangulations of Stokes polytopes. The weights can be assigned a unique numerical value, which makes the sum over Stokes polytopes equal to the $\phi^4$ amplitudes. 

In this paper, we address the issue of the undetermined weights of Stokes polytopes using the generalized BCFW recursion relations. Our primary motivation for using the BCFW recursion relations is to shed some light on the origin of these weights in the calculation of $\phi^4$ amplitudes. We show that the factorization at physical poles along with boundary terms in $\phi^4$ uniquely fixes the weights of the higher-point amplitudes, calculated by summing up over Stokes polytopes, in terms of the weights of the lower-point ones. Further, we see that the non-triviality of these weights is intricately tied to the fact that there are no BCFW recursion relations without boundary terms for $\phi^4$ amplitudes, that completely capture all the factorization channels \cite{12}. This is a consequence of the fact that for two neighbouring shifted legs in tree-level $\phi^4$ Feynman graphs, the residue at infinity is non-vanishing \cite{13}. 

The paper is organized as follows. In section \ref{sec:2.1}, we give an overview of the calculation of amplitudes from the geometry of Stokes polytopes. We briefly review the important notions of quadrangulation, $Q$-compatibility and convex realization of Stokes polytopes. The overview is not exhaustive, and we refer the interested reader to \cite{2,3,4} for complete details. In section \ref{sec:2.2}, we review the generalized BCFW recursion relations for $\phi^4$ amplitudes as given in \cite{9,10}, and discuss the calculation of boundary terms in $\phi^4$ amplitudes. In section \ref{sec:3}, we use the generalized BCFW recursion relations to constrain the weights of higher-point amplitudes in terms of lower-point weights. Firstly, we use the boundary terms to show that the lowest-point weight $\alpha_6$ is fixed uniquely. Then, we calculate the weights of Stokes polytopes corresponding to eight- and ten-point amplitudes and show that these are determined exactly in terms of the six-point weight $\alpha_6$. By substituting the value of $\alpha_6$, we determine the numerical value of eight- and ten-point weights and show that these agree with the results in \cite{3,18}. In section \ref{sec:4}, we generalize our results to an $n$-point amplitude and prove that the weights of the corresponding Stokes polytopes are uniquely fixed by the factorization of the amplitude. We end the paper with a discussion of our results and a brief commentary on using other methods to determine the weights.

Before proceeding we want to add that while this manuscript was being prepared, we came across \cite{18}, which uses `BCFW' like recursion relations as described in \cite{15,16} to determine the weights of the Stokes polytopes in $\phi^4$ and in general, $\phi^p$ theory.

\section{Amplitudes for massless planar $\phi^4$ theory}
In this section, we give an overview of the key results of \cite{3}, where the relationship between planar Feynman graphs in $\phi^4$ theory and positive geometries was established. We focus on the construction of planar massless $\phi^4$ amplitudes at tree-level by summing over the Stokes polytopes.  Further, we also review the key results of \cite{9,10}, where the boundary behaviour of the $\phi^4$ amplitudes was analyzed, and the generalized BCFW recursion relations was presented. Throughout the paper, we have considered the tree-level massless planar $\phi^4$ amplitudes.

\subsection{$\phi^4$ amplitudes from Stokes Polytopes}
\label{sec:2.1}
In \cite{3,4}, the authors extended the seminal work of \cite{1} to show that the planar amplitudes for massless $\phi^4$ theory can be obtained from the positive geometries of a polytope of dimensions $N=\left(\frac{n-4}{2}\right)$ known as Stokes polytopes.

A positive geometry, for example, that of polygons and polytopes, is a closed geometry with boundaries or facets of all co-dimensions. There is a unique meromorphic differential form $\Omega$ that is canonically associated with a positive geometry whose form is fixed by the requirement of having logarithmic singularities at the boundaries and the residue at these is equal to the canonical form of the boundary \cite{11}. These canonical forms link the positive geometries to the scattering amplitudes.

For the analysis of planar amplitudes the planar kinematic variables are used. These variables are defined as 
\begin{equation}
\label{eq:2.1}
X_{i,j}=(P_{ij\ldots j-1})^2\equiv(p_i+p_{i+1}+\cdots +p_{j-1})^2, \qquad 1\leq i<j \leq n ,
\end{equation}
where $p_i$ represents the momentum of the $i$-th particle.
The Mandelstam variables can be expressed in terms of the planar variables as
\begin{equation}
\label{eq:2.2}
s_{ij}=2p_i\cdot p_j= X_{i,j+1}+X_{i+1,j}-X_{i,j}-X_{i+1,j+1}~.
\end{equation}

For the $\phi^4$ amplitudes, the quadrangulations of an $n$-gon, where n is always even, are considered (figure \ref{fig:2}, \ref{fig:3}). The total number of ways to completely quadrangulate an $n=(2I+2)$-gon is equal to the Fuss-Catalan number $F_I=\frac{1}{2I+1}\left({3I \atop I}\right)$. Each quadrangulation of the $n$-gon is associated with a planar Feynman graph with propagators $X_{a_{1}},\ldots, X_{a_{\frac{n-4}{2}}}$. Unlike the planar $\phi^3$ amplitudes\footnote[1]{The massless planar $\phi^3$ amplitudes is obtained from the canonical form associated to the \\ Associahedron~\cite{2}.}, the $\phi^4$ amplitudes cannot be constructed from a unique canonical form related to a positive geometry. In $\phi^4$, for a given number of particles $n=(2I+2)$, there are $F_I$ number of Stokes polytopes whose weighted sum gives the full amplitude. These weights are constrained by the factorization of the amplitudes at the BCFW poles. 

For the construction of Stokes polytopes, it is important to define the notion of compatibility of a quadrangulation with a reference quadrangulation. This follows from the notion of compatibility of a diagonal with the reference quadrangulation and is given in detail in \cite{3,4}. The vertices of a Stokes polytopes with reference quadrangulation $Q$ are the quadrangulations which are compatible with $Q$ (figure \ref{fig:2}). A key result from \cite{3} is that the construction of Stokes polytopes of a given dimension depends on the chosen reference quadrangulation $Q$, and different reference quadrangulations correspond to distinct Stokes polytopes denoted by $\mathcal{S}_n^Q$. 

For a given quadrangulation $Q$ of an $n$-gon, the $Q$-dependent planar scattering form is given as 
\begin{equation}
\label{eq:2.3}
\Omega^Q _n = \sum_{\text{Graphs}}(-1)^{\sigma(\text{flip})}d~ \text{ln}X_{a_{1}}\wedge d~ \text{ln}X_{a_{2}} \ldots \wedge d~ \text{ln}X_{a_{\frac{n-4}{2}}} ~,
\end{equation}
where $\sigma(\text{flip})= \pm 1$. A single quadrangulation does not capture the contributions from all the $\phi^4$ propagators.

With the Stokes polytopes and its respective canonical differential form defined, the pullback of the canonical form on the polytopes gives a form proportional to the partial amplitude corresponding to the $Q$-quadrangulation. To define the pullback, a convex realization of Stokes polytopes is established by imposing a set of constraints\footnote[2]{A positive geometry known as the ABHY associahedra also completely encapsulates the $\phi^4$ amplitudes \cite{4}.}.
This is done by embedding the Stokes polytopes $\mathcal{S}_n^Q$ of dimensions $\left(\frac{n-4}{2}\right)$ inside an Associahedra $\mathcal{A}_n$ of dimensions $(n-3)$. 
\\
For example, in $n=6$ case the constraints corresponding to the reference quadrangulation $Q=(14)$ and the $Q$-compatible quadrangulation $Q_1=(36)$  are given by 
\begin{subequations}
\begin{align}
\label{eq:2.4a}
s_{ij}&=-c_{ij} \qquad \forall~1 \leq i < j \leq n-1 = 5, ~~ |i-j|\geq 2  \\
\label{eq:2.4b}
X_{1,3}&=d_{13}~,~~ X_{1,5}=d_{15}~~~ \text{with}~~ d_{13},~d_{15} > 0~.
\end{align}
\end{subequations}
The constraints in \eqref{eq:2.4a} locate the $3$-D associahedron $\mathcal{A}_6$  inside the kinematic space. The constraints in \eqref{eq:2.4b} locates the $1$-D Stokes polytopes $\mathcal{S}_6^{(14)}$ inside $\mathcal{A}_6$.

The pullback of \eqref{eq:2.3} on the space of $\mathcal{S}_n^Q$ results in 
\begin{equation}
\omega_n^Q =\left( m_n^Q\right)~dX_{a_{1}}\wedge dX_{a_{2}} \ldots \wedge d X_{a_{\frac{n-4}{2}}}~ ,
\end{equation}
where $m_n^Q$ is the rational canonical function associated with the Stokes polytope $\mathcal{S}_n ^Q$ and is given as 
\begin{equation}
m_n^Q = \sum_{g(Q)} \left( \frac{1}{\prod^{\frac{n-4}{2}}_{a=1} X_{i_a,j_a}}\right)~,
\end{equation}
and $g(Q)$ denote tree graphs corresponding to a $Q$-compatible set. 
The weighted sum of these functions over all Stokes polytopes gives the full planar scattering amplitude. 

The computation of $m_n^Q$ is greatly aided by an interesting fact about the $n$-gon, i.e. all its quadrangulations can be determined by a cyclic permutation $\sigma$ of a subset of quadrangulations $\{Q_1,...,Q_I\}$, referred to as the set of primitive quadrangulation. The subset of primitive quadrangulations can be arbitrarily chosen from the set of all qudrangulations for an $n$-gon satisfying the following
\begin{itemize}

\item No two members of this subset are related by cyclic permutations $\sigma$. 
\item All the other quadrangulations can be obtained by cyclic permutations of the quadrangulations of this subset.
\end{itemize}

What this implies is that, for any given $n$, once the rational canonical functions for a given set of primitives $\{Q_1,\ldots ,Q_I \}$ have been calculated, all the other $m_n^Q$ can be computed by a cyclic permutation of the labels of $m_n^{Q}$. The $m_n^{Q}$, where $Q \in \{Q_1,\ldots ,Q_I \}$ are referred to as primitives. Therefore, the master formula for evaluating the amplitudes is given as 
\begin{equation}
\label{eq:2.6}
\widetilde{\mathcal{M}}_n = \sum_{\text{Q}|\text{primitives}}\sum_{\sigma}\alpha_Q~ m_n^{(\sigma\cdot Q)}~,
\end{equation}
where $\alpha_Q$ are positive constants and referred to as the weights. The weights are parametrized only by the primitive quadrangulations, i.e. 
\begin{equation}
\alpha_Q = \alpha_{\overline{Q}}~~ \forall~~\overline{Q} = \sigma \cdot Q~.
\end{equation}
There is a unique choice of $\alpha_Q$'s such that the amplitude determined by the weighted sum over all primitives $\widetilde{\mathcal{M}}_n=\mathcal{M}_n$, where $\mathcal{M}_n$ is the tree-level planar $\phi^4$ amplitude. This is guaranteed if the weights $\alpha_Q$ $\forall~ Q \in \{Q_1, \cdots, Q_I\}$, satisfy the condition that when summed over all canonical forms as in \eqref{eq:2.6}, the residue at each pole of $\widetilde{\mathcal{M}}_n$ is unity\footnote[3]{ It is important to note that the poles referred to here are different from the poles due to a BCFW-shift. The residues referred to here corresponds to the poles $X_{i,j} = 0$ of the meromorphic function $\widetilde{\mathcal{M}}_n(\{X_{i,j}\})$~.} \cite{3,4}.

\subsection{$\phi^4$ amplitudes from generalized BCFW recursion relations}
\label{sec:2.2}
In \cite{10} it was shown that a generalized BCFW recursion relation, which gives a prescription to compute the boundary contributions, can be written for the $\phi^4$ theory.

\begin{figure}[tbp]
\centering % \begin{center}/\end{center} takes some additional vertical space
\includegraphics[width=0.6\textwidth, scale=0.1]{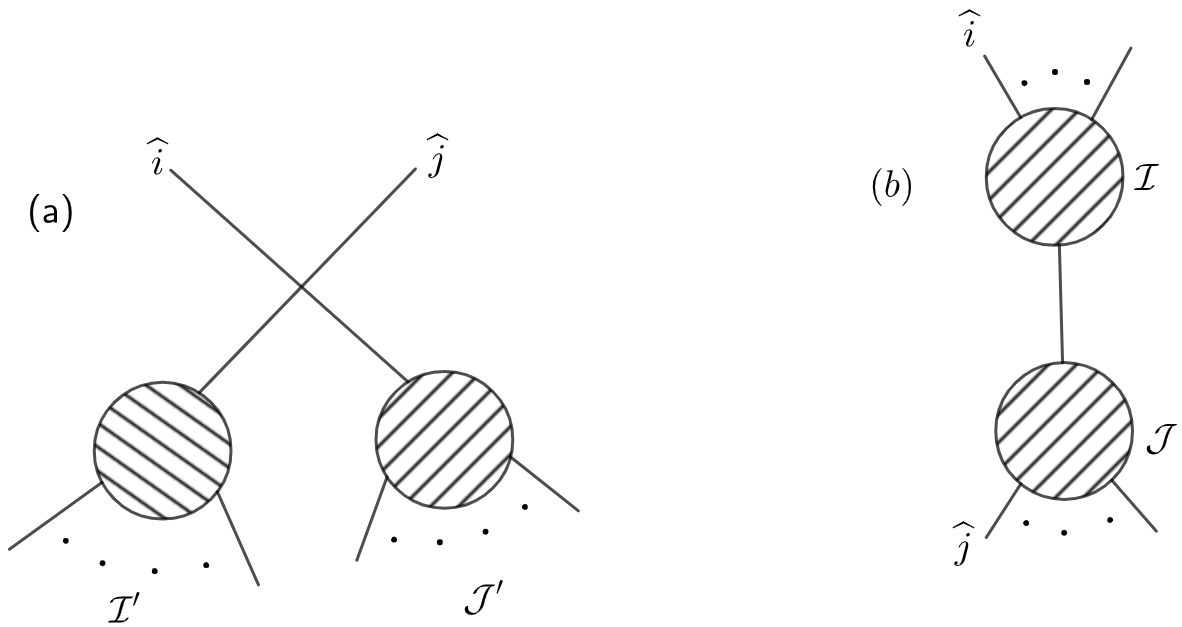}
\caption{\label{fig:1} (a) Diagrams contributing to boundary term $\mathcal{B}_n$. (b) Diagrams contributing to the pole part $\mathcal{P}_n$.}
\end{figure}

Consider the following BCFW shifts denoted as $\langle i|j]$
\begin{equation}
|i \rangle \rightarrow |\hat{i} \rangle = |i \rangle + z |j \rangle, \qquad |j] \rightarrow |\hat{j} ] = |j ] - z |i ].
\end{equation}
Under these shifts the amplitude $\mathcal{M}_n$ becomes a meromorphic function of $z$. There are two categories of Feynman diagram for an $\langle i|j]$-shift. Category (a) is where the particles $i$, $j$ are attached to the same vertex, and category (b) where $i$ and $j$ are attached to different vertices (figure \ref{fig:1}). 

For category (b) of Feynman diagrams, there is at least one propagator on the line connecting $i$ and $j$ that depends linearly on $z$. This gives a factor of $\frac{1}{P^2-z\langle j|P|i]}$ in the expression, which scales as $\sim \frac{1}{z}$. In the limit, $z \to \infty$, such factors have a zero contribution, and therefore category (b) diagrams do not contribute to the boundary terms.

For category (a) diagrams, there is a cancellation of the $z$ terms in the summation of momenta, and therefore it has no $z$ dependence in the expression. In the large $z$ limit, such diagrams have a $\mathcal{O}(z^0)$-behaviour. Therefore, the boundary terms are equal to Feynman diagrams where particles $i$ and $j$ have a common vertex. 

From the above analysis, the generalized on-shell BCFW recursion relations can be given as
\begin{equation}
\mathcal{M}_n = \mathcal{P}_n + \mathcal{B}_n~,
\end{equation}
where $\mathcal{P}_n$ denotes the pole part corresponding to category (b) diagrams and is given as 
\begin{equation}
\label{eq:2.11}
\mathcal{P}_n = \sum_{\mathcal{I}} {\mathcal{M}_L(z_{\mathcal{I}})}~\frac{1}{P^2_{\mathcal{I}}}~{\mathcal{M}_R(z_{\mathcal{I}})}~,
\end{equation}
and $\mathcal{B}_n$ denotes the boundary contributions corresponding to category (a) diagrams and is given as 

\begin{equation}
\label{eq:2.12}
\mathcal{B}_n = \sum_{\mathcal{I}' \cup \mathcal{J}' =\{n\} \setminus \{i,j\}} 
{\mathcal{M}_{\mathcal{I}'}} \frac{1}{P^2_{\mathcal{I}'}}\frac{1}{P^2_{\mathcal{J}'}} {\mathcal{M}_{\mathcal{J}'}}~,
\end{equation}
where $\mathcal{I}'$ and $\mathcal{J}'$ corresponds to all the allowed splitting as in figure \ref{fig:1}.

\section{Determining the weights of Stokes polytopes}
\label{sec:3}

In this section, we use the generalized BCFW recursion relations, as discussed in section \ref{sec:2.2}, to constrain the weights of the Stokes polytopes. Firstly we determine the weights for the six-point amplitudes, whose detailed calculation from the geometry of $\mathcal{S}_6^Q$ is given in \cite{3,4}. Then, we use the six-point amplitudes as input in the generalised recursions for the calculation of eight- and ten-point amplitudes. We match the coefficients of the individual terms appearing in the amplitudes calculated from generalized BCFW with the coefficients of respective terms in the amplitudes calculated from the summation over Stokes polytopes. We show that this puts constraints on the weights that fixes them uniquely.
We use the notation $\alpha_n^{Q}$ to denote the weight for an $n$-point amplitude and corresponding to the primitive quadrangulation $Q$. Also, note that we have referred to the summation over the $Q$-quadrangulations as in \eqref{eq:2.6}, as a summation over the Stokes polytopes $S_n^Q$ throughout the text.

\subsection{Six-point amplitudes}

\begin{figure}[tbp]
\centering % \begin{center}/\end{center} takes some additional vertical space
\includegraphics[width=0.7\textwidth]{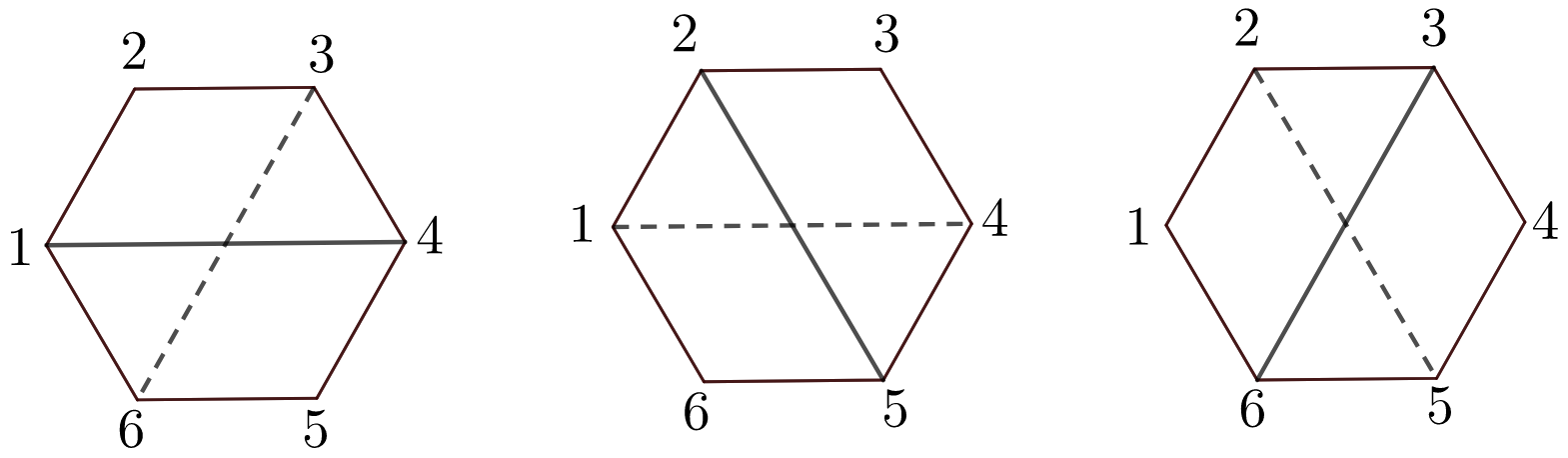}
\caption{\label{fig:2} All possible quadrangulations of a hexagon. The solid-line diagonals represents the reference quadrangulation $Q$ and the dashed-line represents the $Q$-compatible quadrangulation.}
\end{figure}

There are three possible quadrangulations of the $6$-gon as shown in figure \ref{fig:2}. The weighted sum over the Stokes polytopes $\mathcal{S}_6^{Q}$, where $Q \in \{(14),(25),(36)\}$, is given as 
\begin{equation}
\label{eq:3.1}
\widetilde{\mathcal{M}}_6 = \alpha_6^{Q(14)}\left(\frac{1}{X_{1,4}}+\frac{1}{X_{3,6}}\right)+\alpha_6^{Q(25)}\left(\frac{1}{X_{2,5}}+\frac{1}{X_{1,4}}\right)+\alpha_6^{Q(36)}\left(\frac{1}{X_{3,6}}+\frac{1}{X_{2,5}}\right).
\end{equation}

The $n=6$ case has only one primitive, and its cyclic permutation
gives the canonical rational functions corresponding to other quadrangulations, as can be seen from \eqref{eq:3.1}. This implies that the three weights are equal and the six-point amplitude is given as 
\begin{equation}
\label{eq:3.2}
\widetilde{\mathcal{M}}_6 = 2 \alpha_{6}\left(\frac{1}{X_{1,4}}+\frac{1}{X_{2,5}}+\frac{1}{X_{3,6}}\right) = 2 \alpha_{6}\left(\frac{1}{P^2_{123}}+\frac{1}{P^2_{234}}+\frac{1}{P^2_{345}}\right),
\end{equation}
where we used the equation \eqref{eq:2.2} to express the planar variables $X$ in terms of $P^2$. We also dropped the label for quadrangulation in the weight $ \alpha_6$. 

Next, we use the $\langle 1|2]$-shift and apply the BCFW recursion relations. The boundary term has contributions from the diagrams $(123|456)$ and $(612|345)$ and is given as 
\begin{equation}
\label{eq:3.2a}
\begin{split}
\widetilde{\mathcal{B}}^{\langle 1|2]}_6 =~& \widetilde{\mathcal{M}}_4(1,2,3,-P)\frac{1}{P^2_{123}}\widetilde{\mathcal{M}}_4(P,4,5,6)
\\
&+ \widetilde{\mathcal{M}}_4(6,1,2,-P)\frac{1}{P^2_{345}}\widetilde{\mathcal{M}}_4(P,3,4,5)
\\
= & \left(\frac{1}{P^2_{123}}+\frac{1}{P^2_{345}}\right)~.
\end{split}
\end{equation}
The pole part has contribution from the diagram $(561|234)$ and is given as
\begin{equation}
\label{eq:3.2b}
\widetilde{\mathcal{P}}_6^{\langle 1|2]} =  \widetilde{\mathcal{M}}_4(5,6,1,-P)\frac{1}{P^2_{234}}\widetilde{\mathcal{M}}_4(P,2,3,4) = \frac{1}{P^2_{234}}.
\end{equation}
From \eqref{eq:3.2}, \eqref{eq:3.2a}, and \eqref{eq:3.2b} the six-point weight is determined as $\alpha_6=\frac{1}{2}$.

\subsection{Eight-point amplitudes}
\label{sec:3.2}

To determine the eight-point amplitudes, we use the $\langle 1|2]$-shift and then apply BCFW recursion relations. 
The boundary term has three contributions from the diagrams $(123|45678)$, $(12345|678)$ and $(34567|812)$ and is given as
\begin{equation}
\begin{split}
\widetilde{\mathcal{B}}^{\langle 1|2]}_{8}(1,2,...,8) =&~\widetilde{\mathcal{M}}_4(1,2,3,-P)\frac{1}{P^2_{123}} \widetilde{\mathcal{M}}_6(P,4,5,6,7,8)
\\
&+~\widetilde{\mathcal{M}}_4(8,1,2,-P)\frac{1}{P^2_{128}} \widetilde{\mathcal{M}}_6(P,3,4,5,6,7)
\\
&+~\widetilde{\mathcal{M}}_4(3,4,5,-P)\frac{1}{P^2_{345}P^2_{678}}\widetilde{\mathcal{M}}_4(P,6,7,8) ~,
\end{split}
\end{equation}
which on substituting \eqref{eq:3.2} simplifies as 
\begin{equation}
\widetilde{\mathcal{B}}^{\langle 1|2]}_{8} = \frac{2\alpha_{6}}{P^2_{123}} \left(\frac{1}{P^2_{456}} +\frac{1}{P^2_{567}} +\frac{1}{P^2_{678}} \right)+\frac{2\alpha_{6}}{P^2_{128}} \left(\frac{1}{P^2_{345}}+\frac{1}{P^2_{456}} +\frac{1}{P^2_{567}}\right)+\frac{1}{P^2_{345}P^2_{678}} .
\end{equation}
The pole contribution is from two diagrams and is given as 
\begin{equation}
\begin{split}
\label{eq:3.5}
\widetilde{\mathcal{P}}^{\langle 1|2]}_{8}(1,2,...,8)=&~ \widetilde{\mathcal{M}}_{4}(7,8,\widehat{1},-\widehat{P}) \frac{1}{P^2_{178}} \widetilde{\mathcal{M}}_6(\widehat{P},\widehat{2},3,4,5,6) 
\\
&+~ \widetilde{\mathcal{M}}_6(5,6,7,8,\widehat{1},-\widehat{P})\frac{1}{P^2_{234}} \widetilde{\mathcal{M}}_4(\widehat{P},\widehat{2},3,4)
\\
= &~ \frac{2\alpha_{6}}{P^2_{178}} \left(\frac{1}{\widehat{P}^2_{\widehat{2}34}} +\frac{1}{P^2_{345}} +\frac{1}{P^2_{456}} \right)+\frac{2\alpha_{6}}{P^2_{234}} \left(\frac{1}{P^2_{567}}+\frac{1}{P^2_{678}} +\frac{1}{\widehat{P}^2_{\widehat{1}78}}\right)~.
\end{split}
\end{equation}
The above equation can be further simplified by using the following relations
\begin{equation}
\begin{aligned}
\frac{1}{P^2_{178}\widehat{P}^2_{\widehat{2}34}}+\frac{1}{\widehat{P}^2_{\widehat{1}78}P^2_{234}}=\frac{1}{P^2_{178}P^2_{234}}\left(\frac{1}{1-\frac{z_1}{z_2}} +\frac{1}{1-\frac{z_2}{z_1}}\right)= \frac{1}{P^2_{178}P^2_{234}}~ ,
\end{aligned}
\end{equation}
where we used the identity 
\begin{equation}
\begin{aligned}
\label{eq:3.7}
\sum^n_{i=1} \prod^n_{j=1,j\neq i}\frac{1}{1-\frac{z_i}{z_j}}=1 ~.
\end{aligned}
\end{equation}
The $z_i$ and $z_j$ denote the locations of poles.
Using this we get 
\begin{equation}
\begin{aligned}
\widetilde{\mathcal{P}}^{\langle 1|2]}_{8}= 2\alpha_{6} \left(\frac{1}{P^2_{178}P^2_{234}} +\frac{1}{P^2_{178}P^2_{345}} +\frac{1}{P^2_{178}P^2_{456}} + \frac{1}{P^2_{234}P^2_{567}}+\frac{1}{P^2_{234}P^2_{678}} \right)~.
\end{aligned}
\end{equation}
%_____________________________________________________________________________________________%
\begin{figure}[tbp]
\centering % \begin{center}/\end{center} takes some additional vertical space
\includegraphics[width=0.7\textwidth, scale=0.1]{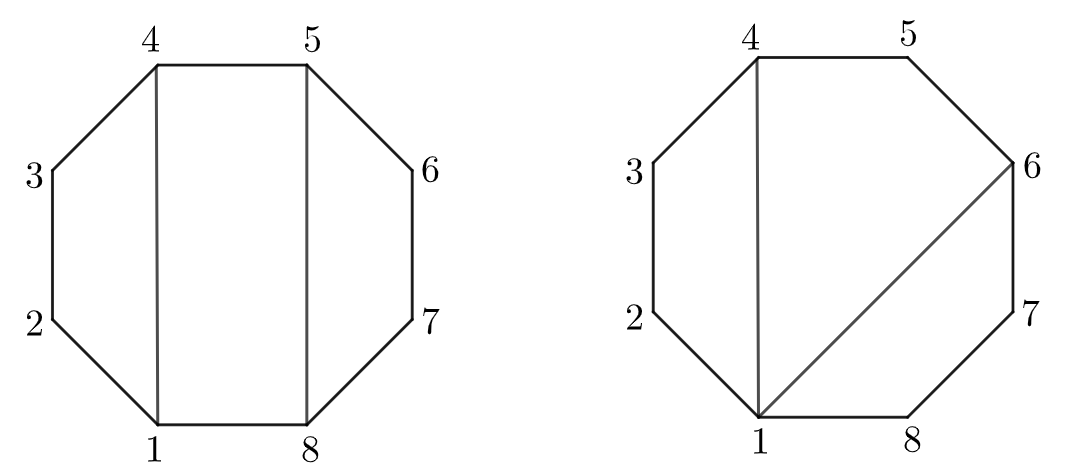}
\caption{\label{fig:3} The two primitives of $n=8$ Stokes polytopes corresponding to quadrangulations $Q=(14,58)$ and $\widetilde{Q}=(14,16)$ respectively.}
\end{figure}
The complete eight-point amplitude, $\widetilde{\mathcal{M}}_8 =\widetilde{\mathcal{P}}_8+\widetilde{\mathcal{B}}_8$, is given as 
\begin{equation}
\label{eq:3.9}
\begin{aligned}
\widetilde{\mathcal{M}}_8 = 2\alpha_{6}& \left(\frac{1}{P^2_{178}P^2_{234}} +\frac{1}{P^2_{178}P^2_{345}} +\frac{1}{P^2_{178}P^2_{456}} + \frac{1}{P^2_{234}P^2_{567}}+\frac{1}{P^2_{234}P^2_{678}} + \frac{1}{P^2_{123}P^2_{456}} \right. 
\\ & 
\left. +\frac{1}{P^2_{123}P^2_{567}}+\frac{1}{P^2_{123}P^2_{678}} + \frac{1}{P^2_{128}P^2_{345}}+\frac{1}{P^2_{128}P^2_{456}} +\frac{1}{P^2_{128}P^2_{567}}\right)+\frac{1}{P^2_{345}P^2_{678}}~.
\end{aligned}
\end{equation}
\\
%_____________________________________________________________________________________________
Now, we calculate the eight-point amplitude by summing up over all the Stokes polytopes $\mathcal{S}^Q_8$. The $n=8$ case has two primitives and in total twelve quadrangulations\footnote[4]{The complete set of quadrangulations for $n=8$ Stokes polytopes are given as \\
$\sigma \cdot Q(14,58) \Rightarrow \mathcal{C}_1=\{(14,58),(25,16),(36,27),(47,38)\}$ and 
\\
$\sigma \cdot \widetilde{Q}(14,16) \Rightarrow \mathcal{C}_2=\{(14,16),(25,27),(36,38),(47,14),(58,25),(16,36),(27,47),(38,58)\}$} (figure \ref{fig:3}), which are the cyclic permutation of labels of the either of the two primitives. The canonical function corresponding to these primitives are given as  
%_____________________________________________________________________________________________
\begin{equation}
\begin{split}
\label{eq-a}
m_8^{Q} &=\left(\frac{1}{X_{1,4}X_{5,8}}+\frac{1}{X_{3,8}X_{4,7}}+\frac{1}{X_{1,4}X_{4,7}}+\frac{1}{X_{3,8}X_{5,8}}\right) \\
& = \left(\frac{1}{P^2_{123}P^2_{567}}+\frac{1}{P^2_{128}P^2_{456}}+\frac{1}{P^2_{123}P^2_{456}}+\frac{1}{P^2_{128}P^2_{567}}\right) \\[5pt] 
m_8^{\widetilde{Q}}&=\left(\frac{1}{X_{1,4}X_{1,6}}+\frac{1}{X_{1,4}X_{5,8}}+\frac{1}{X_{3,6}X_{1,6}}+\frac{1}{X_{3,6}X_{3,8}}+\frac{1}{X_{5,8}X_{3,8}}\right) \\
& = \left(\frac{1}{P^2_{123}P^2_{678}}+\frac{1}{P^2_{123}P^2_{567}}+\frac{1}{P^2_{345}P^2_{678}}+\frac{1}{P^2_{345}P^2_{128}}+\frac{1}{P^2_{567}P^2_{128}}\right) ~.
\end{split}
\end{equation}
\\
Taking the weighted sum over all the $\mathcal{S}_8^Q$ using \eqref{eq:2.6}, and comparing each term to respective term in equation \eqref{eq:3.9}, we get the following relations%_______________________________________________________________________________________
\begin{equation}
\begin{split}
\label{eq:3.11}
2\alpha_8^Q+2\alpha_8^{\widetilde{Q}}= 2\alpha_6 \qquad , \qquad \alpha_{8}^Q+4\alpha_8^{\widetilde{Q}} = 2\alpha_6~.
\end{split}
\end{equation}
%_____________________________________________________________________________________________
Also, the last term in \eqref{eq:3.9} gives the relation 
\begin{equation}
\begin{aligned}
\label{eq:3.12}
\alpha_{8}^Q+4\alpha_8^{\widetilde{Q}}=1~ .
\end{aligned}
\end{equation}
%_____________________________________________________________________________________________
Using \eqref{eq:3.11} and \eqref{eq:3.12},  the weights for eight-point Stokes polytopes is determined to be 
\begin{equation}
\label{eq:3.13a}
\alpha_{8}^Q = \frac{2\alpha_6}{3}=\frac{1}{3}, \qquad \alpha_8^{\widetilde{Q}}=\frac{\alpha_6}{3}=\frac{1}{6}~,
\end{equation}
where the $Q$ and $\widetilde{Q}$ denote the primitive quadrangulations.

\subsection{Ten-point amplitudes}
\label{sec:3.3}

To determine the relation between the weights of $n=10$ Stokes polytopes and $n=6, 8$ weights, we express the $n=8$ amplitude in terms of $\alpha_8^Q$ and $\displaystyle{\alpha_8^{\widetilde{Q}}}$ i.e. as weighted sum over $\mathcal{S}_8^Q$ as 
\begin{equation}
\begin{aligned}
\label{eq:3.13}
\widetilde{\mathcal{M}}_{8}(1,2,...,8) = &~(2\alpha_{8}^Q+2\alpha_{8}^{\widetilde{Q}})\sum_{\sigma\in\mathbb{Z}_8}\left(\frac{1}{P^2_{\sigma(1)\sigma(2)\sigma(3)}P^2_{\sigma(5)\sigma(6)\sigma(7)}}\right) 
 \\ & 
+ (\alpha_{8}^Q+4\alpha_{8}^{\widetilde{Q}})\sum_{\sigma\in\mathbb{Z}_8}\left(\frac{1}{P^2_{\sigma(1)\sigma(2)\sigma(3)}P^2_{\sigma(4)\sigma(5)\sigma(6)}}\right) .
\end{aligned}
\end{equation}

Again, we use the $\langle 1|2]$-shift and apply generalized BCFW recursion relations. The boundary terms has four contributions form the diagrams $(23|456789(10))$, $(345|6789(10))$, 
\\ $(34567|89(10))$ and $(3456789|10)$ and is given as 
%_____________________________________________________________________________________________
\begin{equation}
\begin{aligned}
\label{eq:3.14}
\widetilde{\mathcal{B}}^{\langle 1|2]}_{10}(1,2,...,10) = &~ \widetilde{\mathcal{M}}_{4}(1,2,3,-P)\frac{1}{P^2_{123}}\widetilde{\mathcal{M}}_{8}(P,4,...10) 
 \\ &
+ \widetilde{\mathcal{M}}_{4}(1,2,-P_1,-P_2)
\left(\frac{1}{P^2_{345}}\widetilde{\mathcal{M}}_4(P_1,3,4,5)\right)
\left(\frac{1}{P^2_{12345}}\widetilde{\mathcal{M}}_{6} (P_2,6,...,10)\right)
\\
&
+ \widetilde{\mathcal{M}}_4(1,2,-P_1,-P_2)\left(\frac{1}{P^2_{345}}\widetilde{\mathcal{M}}_6(P_1,3,..,7)\right)
\left(\frac{1}{P^2_{12345}}\widetilde{\mathcal{M}}_{4} (P_2,8,9,10)\right) 
\\
&
+ \widetilde{\mathcal{M}}_4(10,1,2,-P) \frac{1}{P^2_{12(10)}}\widetilde{\mathcal{M}}_8(P,3,..,9) ~.
\end{aligned}
\end{equation}
Using \eqref{eq:3.2} and \eqref{eq:3.13} the boundary term is simplified to 
%_____________________________________________________________________________________________%
\begin{equation}
\begin{aligned} 
\widetilde{\mathcal{B}}^{\langle 1|2]}_{10} = &~ \frac{1}{P_{123}^2} \left[ \left(2 \alpha _{8}^Q+2 \alpha_{8}^{\widetilde{Q}}\right) \left(\frac{1}{P_{567}^2
   P_{45678}^2}+\frac{1}{P_{678}^2
   P_{56789}^2}+\frac{1}{P_{789}^2
   P_{6789(10)}^2}+\frac{1}{P_{456}^2 P_{89(10)}^2}\right) \right.
\\
&
   \left. +\left(\alpha_{8}^Q+4 \alpha_{8}^{\widetilde{Q}}\right)\left(\frac{1}{P_{567}^2P_{89(10)}^2} 
   + \frac{1}{P_{456}^2
   P_{45678}^2}+\frac{1}{P_{678}^2 P_{45678}^2}+\frac{1}{P_{567}^2 P_{56789}^2}+\frac{1}{P_{678}^2 P_{6789(10)}^2} \right. \right.
\\
&
   \left. \left.
   + \frac{1}{P_{89(10)}^2 P_{6789(10)}^2}+\frac{1}{P_{456}^2 P_{789}^2}
   +\frac{1}{P_{56789}^2 P_{789}^2}\right)\right]
   +\frac{2\alpha _6}{P_{345}^2 P_{12345}^2}
   \left(\frac{1}{P_{789}^2}+\frac{1}{P_{89(10)}^2}+\frac{1}{P_{678}^2}\right)
\\
&
+\frac{2 \alpha_6}{P_{89(10)}^2 P_{34567}^2}\left(\frac{1}{P_{456}^2}+\frac{1}{P_{567}^2}+\frac{1}{P_{345}^2}\right)
+\frac{1}{P_{12(10)}^2}\left[ \left (2\alpha_{8}^Q+2 \alpha_{8}^{\widetilde{Q}}\right)\left(\frac{1}{P_{456}^2 P_{34567}^2}+\frac{1}{P_{567}^2 P_{45678}^2} \right. \right.
\\
&
\left. \left. +\frac{1}{P_{678}^2 P_{56789}^2}+\frac{1}{P_{345}^2 P_{789}^2}\right) +\left(\alpha_{8}^Q+4 \alpha_{8}^{\widetilde{Q}}\right)
 \left(\frac{1}{P_{456}^2P_{789}^2}  +\frac{1}{P_{345}^2 P_{34567}^2}+\frac{1}{P_{567}^2 P_{34567}^2} \right. \right.
 \\
 &
 \left. \left.
 +\frac{1}{P_{456}^2 P_{45678}^2}+\frac{1}{P_{567}^2 P_{56789}^2}+\frac{1}{P_{789}^2 P_{56789}^2}+\frac{1}{P_{345}^2 P_{678}^2}+\frac{1}{P_{45678}^2 P_{678}^2}\right)\right] .
\end{aligned}   
\end{equation}
%_____________________________________________________________________________________________%
The pole part is given as 
%_____________________________________________________________________________________________%
\begin{equation}
\begin{aligned}
\label{eq:3.16}
\widetilde{\mathcal{P}}^{\langle 1|2]}_{10}(1,2,...,10) =&~ \widetilde{\mathcal{M}}_{4}(9,10,\widehat{1},-\widehat{P})\frac{1}{P^2_{19(10)}}\widetilde{\mathcal{M}}_{8}(\widehat{P},\widehat{2},3,...,8) 
\\
& 
+ \widetilde{\mathcal{M}}_{6}(7,8,9,10,\widehat{1},-\widehat{P})
\frac{1}{P^2_{23456}}\widetilde{\mathcal{M}}_6(\widehat{P},\widehat{2},3,4,5,6)
\\
& 
+ \widetilde{\mathcal{M}}_8(5,6,7,8,9,10,\widehat{1},-\widehat{P})\frac{1}{P^2_{234}}\widetilde{\mathcal{M}}_4(\widehat{P},\widehat{2},3,4)~,
\end{aligned}
\end{equation}
%_____________________________________________________________________________________________%
which is expanded as 
\begin{equation}
\begin{aligned}
\label{eq:3.17}
\widetilde{\mathcal{P}}^{\langle 1|2]}_{10} =
&~ \frac{1}{P_{19(10)}^2} \left[ \left(2 \alpha _{8}^Q+2\alpha_{8}^{\widetilde{Q}}\right) \left(\frac{1}{P_{456}^2 P_{34567}^2}+\frac{1}{P_{567}^2 P_{45678}^2}+\frac{1}{P_{345}^2 P_{\widehat{2}3456}^2}+\frac{1}{P_{678}^2 P_{\widehat{2}34}^2}\right) \right. 
\\ 
& 
\left.
+\left(\alpha _{8}^Q+4 \alpha_{8}^{\widetilde{Q}}\right) \left(\frac{1}{P_{567}^2 P_{\widehat{2}34}^2}+\frac{1}{P_{345}^2 P_{34567}^2}+\frac{1}{P_{567}^2 P_{34567}^2}+\frac{1}{P_{456}^2 P_{45678}^2}+\frac{1}{P_{456}^2 P_{\widehat{2}3456}^2} \right. \right.
\\
&   
\left. \left.
   +\frac{1}{P_{\widehat{2}34}^2 P_{\widehat{2}3456}^2}+\frac{1}{P_{345}^2 P_{678}^2}+\frac{1}{P_{45678}^2 P_{678}^2}\right)\right] +\frac{4\alpha _6^2}{P_{23456}^2} \left(\frac{1}{P_{89(10)}^2}+\frac{1}{P_{9(10)\widehat{1}}^2}+\frac{1}{P_{789}^2}\right)
\\
&
~~ \times \left(\frac{1}{P_{456}^2}+\frac{1}{P_{\widehat{2}34}^2}+\frac{1}{P_{345}^2}\right)
+\frac{1}{P_{234}^2}\left[ \left(2 \alpha _{8}^Q+2 \alpha_{8}^{\widetilde{Q}}\right) \left(\frac{1}{P_{678}^2 P_{56789}^2}+\frac{1}{P_{789}^2 P_{6789(10)}^2} \right. \right.
\\
&
\left. \left.
+\frac{1}{P_{89(10)}^2 P_{\widehat{1}789(10)}^2}+\frac{1}{P_{567}^2 P_{9(10)\widehat{1}}^2}\right)+\left(\alpha _{8}^Q+4 \alpha_{8}^{\widetilde{Q}}\right)\left(\frac{1}{P_{678}^2 P_{9(10)\widehat{1}}^2}+\frac{1}{P_{567}^2 P_{56789}^2} +\frac{1}{P_{789}^2 P_{56789}^2}
\right. \right.
\\
&
\left. \left.
+\frac{1}{P_{678}^2 P_{6789(10)}^2}+\frac{1}{P_{789}^2 P_{\widehat{1}789(10)}^2}+\frac{1}{P_{9(10)\widehat{1}}^2 P_{\widehat{1}789(10)}^2}+\frac{1}{P_{567}^2 P_{89(10)}^2}+\frac{1}{P_{6789(10)}^2 P_{89(10)}^2}\right)\right].
\end{aligned}
\end{equation}
%_____________________________________________________________________________________________%
Now, making repeated use of the identity \eqref{eq:3.7}, the above equation can be further simplified to
\begin{equation}
\begin{aligned}
\widetilde{\mathcal{P}}^{\langle 1|2]}_{10} = &~ \frac{1}{P_{19(10)}^2} \left[ \left(2 \alpha _{8}^Q+2  \alpha_{8}^{\widetilde{Q}}\right) \left(\frac{1}{P_{456}^2 P_{34567}^2}+\frac{1}{P_{567}^2 P_{45678}^2}+\frac{1}{P_{345}^2 P_{23456}^2}+\frac{1}{P_{678}^2 P_{234}^2}\right) \right. 
\\ 
& \left.
+~\left(\alpha _{8}^Q+4 \alpha_{8}^{\widetilde{Q}}\right) \left(\frac{1}{P_{567}^2 P_{234}^2}+\frac{1}{P_{345}^2 P_{34567}^2}+\frac{1}{P_{567}^2 P_{34567}^2}+\frac{1}{P_{456}^2 P_{45678}^2}+\frac{1}{P_{456}^2 P_{23456}^2} \right. \right.
\\
&
\left. \left.
+\frac{1}{P_{234}^2 P_{23456}^2}+\frac{1}{P_{345}^2 P_{678}^2}+\frac{1}{P_{45678}^2 P_{678}^2}\right)\right]
+\frac{4\alpha _6^2}{P_{23456}^2} \left(\frac{1}{P_{89(10)}^2}+\frac{1}{P_{789}^2}\right) \left(\frac{1}{P_{456}^2}+\frac{1}{P_{345}^2}\right)
\\
& 
 +\frac{1}{P_{234}^2}\left[\left( 2 \alpha _{8}^Q+2  \alpha_{8}^{\widetilde{Q}} \right)\left(\frac{1}{P_{678}^2 P_{56789}^2}+ \frac{1}{P_{789}^2 P_{6789(10)}^2} +\frac{1}{P_{89(10)}^2 P_{23456}^2}\right) \right.  +\left(\alpha _{8}^Q+4  \alpha_{8}^{\widetilde{Q}}\right)
\\
&
\left.
~\times \left(\frac{1}{P_{567}^2 P_{56789}^2}+\frac{1}{P_{789}^2 P_{56789}^2}+\frac{1}{P_{678}^2 P_{6789(10)}^2}+\frac{1}{P_{567}^2 P_{89(10)}^2}+\frac{1}{P_{789}^2 P_{23456}^2}+\frac{1}{P_{6789(10)}^2 P_{89(10)}^2}\right)\right],
\end{aligned}
\end{equation}
%_____________________________________________________________________________________________%
\\
where we have used the relations in \eqref{eq:3.11}, \eqref{eq:3.12}, and the fact that  $2\alpha_{6} = 4 \alpha^2_6$ , since $\alpha_6=\frac{1}{2}$, in making the simplifications.

Now, for the $n=10$ case there are seven primitive Stokes polytopes and in total fifty-five quadrangulations. These correspond to Cube type, Snake type, Lucas type and Mixed type Stokes polytopes \cite{3}. Taking the weighted sum over all the Stokes polytope $\mathcal{S}_{10}^Q$ and comparing to the amplitude $\displaystyle{\widetilde{\mathcal{M}}_{10}=\widetilde{\mathcal{P}}^{\langle 1|2]}_{10}+\widetilde{\mathcal{B}}^{\langle 1|2]}_{10}}$~ computed by BCFW recursions, we get the following set of equations that constrain the ten-point weights as
\\
%_____________________________________________________________________________________________%
\begin{equation}
\begin{split} 
\label{eq:3.19}
4\alpha_{10}^{Q_b}+2\alpha_{10}^{Q_c}+2\alpha_{10}^{Q_d}+2\alpha_{10}^{Q_e}+2\alpha_{10}^{Q_f}+2 \alpha_{10}^{Q_g}  &= 2\alpha_6 ~,
\\[4pt]
2\alpha_{10}^{Q_a}+\alpha_{10}^{Q_c}+\alpha_{10}^{Q_d}+2\alpha_{10}^{Q_f}+2\alpha_{10}^{Q_g} &= 2\alpha_{8}^Q+2\alpha_{8}^{\widetilde{Q}}~,
\\[4pt]
\alpha_{10}^{Q_a}+2\alpha_{10}^{Q_b}+  \alpha_{10}^{Q_c}+2\alpha_{10}^{Q_e}+2\alpha_{10}^{Q_f}+2\alpha_{10}^{Q_g}  &= 2\alpha_6 ~,
\\[4pt]
2\alpha_{10}^{Q_b}+\alpha_{10}^{Q_c}+\alpha_{10}^{Q_d}+4\alpha_{10}^{Q_e}+2\alpha_{10}^{Q_f}+2\alpha_{10}^{Q_g}  &= 2\alpha_{8}^Q+2\alpha_{8}^{\widetilde{Q}} ~,
\\[4pt] \alpha_{10}^{Q_a}+4\alpha_{10}^{Q_b}+3\alpha_{10}^{Q_c}+2\alpha_{10}^{Q_d}+2\alpha_{10}^{Q_e}+2\alpha_{10}^{Q_f}  &= 2\alpha_6 ~,
\\[4pt]
\alpha_{10}^{Q_a}+2\alpha_{10}^{Q_b}+\alpha_{10}^{Q_d}+2\alpha_{10}^{Q_e}+2\alpha_{10}^{Q_f}+2\alpha_{10}^{Q_g}  &= 4\alpha_6^2 ~,
\\[4pt] 
\alpha_{10}^{Q_a}+4\alpha_{10}^{Q_b}+2\alpha_{10}^{Q_c}+3\alpha_{10}^{Q_d}+2\alpha_{10}^{Q_e}+2\alpha_{10}^{Q_g}  &= \alpha_{8}^Q+4\alpha_{8}^{\widetilde{Q}} ~
\end{split}
\end{equation}
where $\{Q_a, Q_b, Q_c, Q_d, Q_e, Q_f, Q_g\}$ correspond to set of quadrangulations that form the primitives of $n=10$ Stokes polytopes.

Substituting \eqref{eq:3.11} in the above equation and solving for the seven undetermined $\alpha_{10}$'s in terms of $\alpha_6$ we get 
%_____________________________________________________________________________________________%
\begin{equation}
\begin{aligned} 
\label{eq:3.20}
\alpha_{10}^{Q_a} &= \frac{1}{12} (12 \alpha_{6}^2 - \alpha_{6}) = \frac{5}{24}~~~~, 
\qquad
\alpha_{10}^{Q_b} = \frac{1}{12} (12 \alpha_{6}^2 - 5 \alpha_{6}) = \frac{1}{24}~, 
\\[4pt]
\alpha_{10}^{Q_c} &= \frac{1}{12} (19\alpha_{6} - 36 \alpha_{6}^2) = \frac{1}{24}~,
\qquad
\alpha_{10}^{Q_d} = \frac{1}{12} (12\alpha_{6}^2-5\alpha_{6})= \frac{1}{24} ~,
\\[4pt]
\alpha_{10}^{Q_e} &= \frac{\alpha_6}{6} = \frac{1}{12} \qquad \qquad \qquad  ~,
\qquad
\alpha_{10}^{Q_f} = \frac{1}{4} (4\alpha_6^2-\alpha_6)= \frac{1}{8}~,
\\[4pt]
\alpha_{10}^{Q_g} &= \frac{1}{4}(3 \alpha_6 - 4\alpha^2_6) =\frac{1}{8} \qquad ,
\end{aligned} 
\end{equation}
where we substituted $\alpha_6=\frac{1}{2}$ in the end. The weights determined in \eqref{eq:3.13a} and \eqref{eq:3.20} are in perfect agreement with the results in \cite{3,15}.

\section{Generalization to higher-point amplitudes}
\label{sec:4}

\subsection{Overview of the proof}
\label{sec:4.1}
Firstly, it is useful to introduce the following notations. Let $\mathcal{Q}_n$ denote the complete set of primitive quadrangulations, i.e. the set $\{Q_1, \cdots Q_I \}$, of an $n$-gon. Let $\alpha_{n}^{Q}$ be the set of weights corresponding to the primitive quadrangulations $Q \in \mathcal{Q}_n$, and let $\widetilde{\alpha}_6= 2 \alpha_6$. Then we would like to prove that if the $\langle i|j]$-shift uniquely fixes the weights for $(n-2)$-point amplitude such that $\widetilde{\mathcal{M}}_{n-2}=\mathcal{M}_{n-2}$, then the weights for the $n$-point amplitude are also uniquely fixed such that $\widetilde{\mathcal{M}}_{n} =\mathcal{M}_{n}$.
To prove this statement it is sufficient to prove that, in general for an $n$-point planar $\phi^4$ amplitude, the weights obey the condition that when summed over all canonical forms as in \eqref{eq:2.6}, the residue at each pole of $\widetilde{\mathcal{M}}_n$ (i.e. residue at poles  $X_{i,j} = 0$) is unity. Since each $X_{i,j}=0$ corresponds to  simple poles, the residues of $\left. \widetilde{\mathcal{M}}_n \right| _{X_{i,j}=0}$ are simply the coefficients of the terms of the form
\begin{equation}
Y=\frac{1}{X_1 X_2 \cdots X_{\frac{n-4}{2}}}~,
\end{equation}
where one of $X = X_{i,j}$, in equation \eqref{eq:2.6}. 

Without any loss of generality, we chose the $\langle 1|2]$-shift to prove our statement. Also for the purpose of the proof, it is important to determine the scaling of amplitudes with respect to the six-point weight $\widetilde{\alpha}_6$. Consider the six-point amplitude $\widetilde{\mathcal{M}}_6$, which is $ \propto \widetilde{\alpha}_6$ as can be seen in \eqref{eq:3.2}. For an $n$-point amplitude, the boundary part and the pole part are constructed recursively from lower-point amplitudes as in \eqref{eq:2.11} and \eqref{eq:2.12}. Since we take the six-point amplitude as the input in these recursion relations, each product of lower-point amplitudes appearing in the factorization of the $n$-point amplitude must be proportional to some power of $\widetilde{\alpha}_6$ \footnote[5]{It does not matter which power of $\widetilde{\alpha}_6$ because $\widetilde{\alpha}_6$ raised to any power is equal to one, since $\widetilde{\alpha}_6=1$.}. For example, the eight-point amplitude obey a factorization which is schematically given as $\widetilde{\mathcal{M}}_8 \sim \widetilde{\mathcal{M}}_4 \times \widetilde{\mathcal{M}}_6$ and is therefore $\propto \widetilde{\alpha}_6$.
The ten-point amplitude has the following factorization, schematically given as 
\begin{equation}
\widetilde{\mathcal{M}}_{10} \sim \widetilde{\mathcal{M}}_{4} \times \widetilde{\mathcal{M}}_{8} + \widetilde{\mathcal{M}}_{6} \times \widetilde{\mathcal{M}}_{6}~,
\end{equation}
where the first term is $\propto \widetilde{\alpha}_6$ and the second terms is $\propto (\widetilde{\alpha}_6)^2$.

Let us assume that the above statement is true for a $k$-point amplitude, where $k=12,14, \cdots ,(n-2)$. That is, each product of lower-point amplitudes appearing in the factorization of $\widetilde{\mathcal{M}}_{k}$ , is proportional to some power of $\widetilde{\alpha}_6$. This implies that the residues at the poles $(X_{i,j} = 0)$ of $\widetilde{\mathcal{M}}_{k}$ are equal to unity, since $\widetilde{\alpha}_6=1$.

We prove the above statement for the $n$-point amplitude in section \ref{sec:4.2}. The proof follows from the following steps
\begin{itemize}
\item We use the $\langle 1|2]$-shift to obtain the correct factorization of the $n$-point amplitude  $\widetilde{\mathcal{M}}_n$. 
\item Using \eqref{eq:2.6} we express a general $l$-point amplitude, $\widetilde{\mathcal{M}}_l$ for $l=6,8, \cdots, n$~,  as the weighted sum over the corresponding Stokes polytopes $\mathcal{S}_l^Q$.
\item We show by induction that each term appearing in the factorization of $\widetilde{\mathcal{M}}_n$ can only be proportional to some power of $\widetilde{\alpha}_6$, and hence is equal to one. 
\item Using the above, we show that this puts the condition on the weights for the $n$-point amplitude that the residues on the poles $X_{i,j}=0$ of $\widetilde{\mathcal{M}}_n$ have a unit contribution.
\end{itemize}

\subsection{$n$-point amplitudes}
\label{sec:4.2}

Consider the above statement for $n=8$ and $n=10$.
The statement holds true for these as can be seen from equations \eqref{eq:3.11} and \eqref{eq:3.19}. The statement holds true for the $k = 12,14, \cdots (n-2)$ -point amplitude by our assumption.

Now, we prove that this statement is true for the $n$-point amplitude. 
The $n$-point amplitude, determined from the $\langle 1|2]$-shift is given as 
\begin{equation}
\label{eq:4.2}
\widetilde{\mathcal{M}}_{n}^{\langle 1|2]} = \widetilde{\mathcal{P}}_n + \widetilde{\mathcal{B}}_n~.
\end{equation}
The $\widetilde{\mathcal{P}}_n$ and $\widetilde{\mathcal{B}}_n$ are constructed recursively from the lower-point amplitudes as given in \eqref{eq:2.11} and \eqref{eq:2.12}, respectively. Therefore, it can be easily seen that the full amplitude obeys the following factorization scheme
\begin{equation}
\begin{split}
\label{eq:4.3}
\widetilde{\mathcal{M}}_{n}^{\langle 1|2]} \sim \widetilde{\mathcal{M}}_{4} \times \widetilde{\mathcal{M}}_{n-2}+ \widetilde{\mathcal{M}}_{6} \times \widetilde{\mathcal{M}}_{n-4} + \ldots + \widetilde{\mathcal{M}}_{q} \times \widetilde{\mathcal{M}}_{n-q+2}~,
\end{split}
\end{equation}
where 
\begin{equation}
\label{eq:4.4}
q = \Bigg\{  {~\frac{n}{2} \qquad ~~~ \text{if}~ \frac{n}{2}~ \text{is even} \atop \left(\frac{n}{2}+1\right)~~ \text{if}~ \frac{n}{2}~ \text{is odd .}}~
\end{equation}
Every term in \eqref{eq:4.3} has an additional factor of $\left(\frac{1}{P^2}\right)^s$, where $s=1$ if the term is form the factorization of $\widetilde{\mathcal{P}}_n$, and $s \geq 1$ if the term is from $\widetilde{\mathcal{B}}_n$.
Substituting \eqref{eq:2.6} for each $\widetilde{\mathcal{M}}$ in the factorization in \eqref{eq:4.3} we get 
%_________________
\begin{equation}
\begin{split}
\label{eq:4.5}
\widetilde{\mathcal{M}}_{n}^{\langle 1|2]} \sim \sum^{q}_{l=4,6, \ldots} \left(\sum_{Q_L,\sigma}  \alpha_{l}^{Q_L} m_l^{(\sigma \cdot Q_L)} \right) \times \left( \sum_{Q_R,\sigma'} \alpha_{n-l+2}^{Q_R}  m_{n-l+2}^{\left(\sigma' \cdot Q_R \right)} \right)~,
\end{split}
\end{equation}
where $Q_L \in \mathcal{Q}_l$ and $Q_R \in \mathcal{Q}_{n-l+2}$~. 
The $l=4$ case in \eqref{eq:4.5} corresponds to the four-point amplitude $\widetilde{\mathcal{M}}_4$, for which $m_4 = \pm 1$ and the weight $\alpha_4=1$.

Under the $\langle 1|2]$-shift, the amplitude $\widetilde{\mathcal{M}}^{\langle 1|2]}_n$ has two types of terms. Terms of type (A) do not depend on the shifted variables $| \hat{1} \rangle$ and $| \hat{2} ]$. Terms of type (B) are functions of these shifted variables. For terms of type (B), repeated use of the identity \eqref{eq:3.7} removes the dependence on the shifted variables. 

The left-hand side of \eqref{eq:4.5} is equal to the sum over $\mathcal{S}_n^Q$ and is given as 
\begin{equation}
\begin{split}
\label{eq:4.6}
\widetilde{\mathcal{M}}_n = \sum_{Q} \sum_{\sigma} \alpha_n^{Q} ~m_n^{(\sigma \cdot Q)}~,
\end{split}
\end{equation}
where $Q \in \mathcal{Q}_n$. After summing up over all rational canonical functions, we compare the residue corresponding to the pole $X_{i,j}=0$ 
in equations \eqref{eq:4.5} and \eqref{eq:4.6} \footnote[6]{This is done after reinserting the correct pre-factor of $\left(\frac{1}{P^2}\right)^s$ in the respective terms.}. This gives the following constraint on the weights
\begin{equation}
\label{eq:4.8}
\sum_{Q} \mathcal{C}_{Q}~ \alpha_n^{Q} = \left(\sum_{Q_L} \mathcal{C}_{Q_L}~\alpha_l^{Q_L}\right) \left( \sum_{Q_R}\mathcal{C}_{Q_R}  \alpha_{n-l+2}^{Q_R} \right)~,
\end{equation}
for some $l \in \{4,6,...,q\}$, where $q$ is defined in \eqref{eq:4.3}, and $\mathcal{C}_{Q}$, $\mathcal{C}_{Q_L}$ and $\mathcal{C}_{Q_R}$ are positive constants. The $\mathcal{C}_{Q}$'s count the number of times 
the weight $\alpha_l^{Q}$ appear in the residue.

Now, from the factorization of $\widetilde{\mathcal{M}}_{n}$ as in \eqref{eq:4.3}, we have that the residues at $X_{i,j}=0$ as in the right hand side of \eqref{eq:4.8}, can only be equal to some power of $\widetilde{\alpha}_6$. This follows from the fact that $\widetilde{\mathcal{M}}_6,~\widetilde{\mathcal{M}}_8  \propto \widetilde{\alpha}_6$, and from our assumption that the products of lower-point amplitudes in the factorization of $\widetilde{\mathcal{M}}_{k}$, where $k=10, 14, \cdots, (n-2)$, is proportional to a power of $\widetilde{\alpha}_6$. Therefore, the product of the lower-point amplitudes in \eqref{eq:4.3} can only be proportional to some power of $\widetilde{\alpha}_6$, which is equal to unity. This implies that the right hand side of \eqref{eq:4.8} is equal to one.

The same procedure can be followed for any arbitrary pole $X=0$ of $\widetilde{\mathcal{M}}_n$, which will give an equation of the form \eqref{eq:4.8}. This implies that for any $n$-point planar $\phi^4$ amplitude, the corresponding weights obey the constraint that 
when summed over all canonical forms as in \eqref{eq:2.6}, each and every pole $X_{i,j}=0$ of the meromorphic function $\widetilde{\mathcal{M}}_n=\widetilde{\mathcal{M}}_n\left(\{X_{i,j}\}\right)$ has a unit residue. \\
This completes our proof.  \hspace{9.5cm} \emph{Q.E.D.}

\section{Summary and Discussion}

The geometric formulation of scattering amplitudes is opening up new ways of thinking about QFTs. The results have been striking in the supersymmetric theories such as the $\mathcal{N}=4$ SYM, where the geometry of the polytope referred to as the Amplituhedron, completely encapsulates the amplitudes at all orders \cite{1}. An excellent understanding of amplitude at tree-level in non-supersymmetric theories such as the scalar massless planar $\phi^3$, $\phi^4$ and in general, $\phi^p$ theories has been propelled by the work in \cite{2,3,4,5}. However, it was shown in \cite{3} that there is no single polytope structure for a given dimension that completely encapsulates the $\phi^4$ amplitudes at tree-level. Instead, there is a family of Stokes polytopes, whose weighted sum gives the complete $\phi^4$ amplitude.

In this paper, we addressed the issue of computing the weights. We showed that the factorization of the $\phi^4$ amplitudes at the `BCFW' poles put strong constraints on the weights. We showed that the boundary terms of $n=6$ amplitudes uniquely fixed the value of the lowest-point weight as $\alpha_6=\frac{1}{2}$.   
Further, we explicitly calculated the weights for $n=8$ and $n=10$ cases in section \ref{sec:3.2} and \ref{sec:3.3} respectively, and showed that these are determined in terms of the six-point weight $\alpha_6$. In section \ref{sec:4}, we generalized our result to higher-point amplitudes. 
Using the generalized BCFW recursions, we proved that for any given $n$, the weights obey the condition that when summed over all canonical forms, the residue at each pole is unity. This condition fixes the values of the weights uniquely for which the sum of the canonical forms is equal to the $\phi^4$ scattering amplitudes.

A key feature of our analysis of the weights relied on the boundary terms of $\phi^4$ amplitudes, which correspond to the $\mathcal{O}(z^0)$-behaviour of the amplitudes at large $z$. In general, the full $n$-point tree amplitudes in $\phi^4$ theory cannot be recovered solely from the information of its singularities, which correspond to its residues on the factorization channel \cite{12}. There are no BCFW recursion relations without boundary terms in $\phi^4$, and hence for fixing the weights of the Stokes polytopes, the boundary terms play a crucial role. Also, we know that a single positive geometry is not sufficient to capture the full $\phi^4$ amplitudes and only a properly weighted sum, where the weights are fixed by the boundary terms, is required to capture all the factorization channels. In contrast, in the $\phi^3$ theory, the residue at infinity is absent, and the tree-level amplitudes can be fully determined from a single positive geometry of the Associahedron. This is guaranteed by a special property of the $\phi^3$ amplitudes, which follows from the ``projectivity'' of the canonical form for $\phi^3$ \cite{15}. This also holds true for the tree-level $\mathcal{N}=4$ SYM amplitudes, which have a $\mathcal{A}^{\mathcal{N}=4~ \text{SYM}}_n \sim \frac{1}{z^s}$ fall-off, where $s \geq 1$, under a super-BCFW shift. The residue at $\infty$ is vanishing in this theory, and its amplitudes can be obtained solely by a BCFW-type recursion relations \cite{12,13}. Further, a single positive geometry, that of the Amplituhedron, is sufficient to capture all the tree-level amplitudes in $\mathcal{N}=4$ SYM. 

An important question that arises from our analysis is, 
whether the absence of the residue at infinity in a given QFT, guarantees that the full tree-level amplitudes of the theory, which captures all the factorization channels, can be determined from a single positive geometry corresponding to a specific polytope in the kinematic subspace? As we saw from our analysis that this is not the case for the tree-level $\phi^4$ amplitudes, and hence it required a weighted sum over positive geometries. It would also be interesting to investigate this for polynomial interactions of type $\lambda_3 \phi^3 + \lambda_4 \phi^4$, which are determined by a weighted sum over certain positive geometries known as the Accordiohedron \cite{5,14}.

A shortcoming of our analysis was that an explicit formula for the higher-point weights is still missing. One of the limiting factors was that there does not exist a general method to count the number of primitive Stokes polytopes for dimensions higher than three. However, despite this limitation, we could make a general statement about the weights in our proof in section \ref{sec:4}. Also, the computation of $\alpha_n^Q$ relied on the correct factorization of the amplitudes at the `BCFW' poles, $\widehat{P}=0$. This is a step back from the Amplituhedron program, where the geometry of the polytopes is sufficient to completely determine the amplitudes.
Further, we believe that the extension of the `BCFW'-type recursion relations for $\phi^3$ amplitudes, as presented in \cite{15,16,17}, to $\phi^4$ amplitudes \cite{18} can help answer the questions posed above. We plan to address these questions in our future work.
\\[5pt]
\\
\textbf{Acknowledgements}
\\
\\
We are extremely thankful to Alok Laddha for suggesting the problem, for numerously stimulating discussions on the topic and comments on the draft. We are also thankful to Dileep Jatkar and PB Aneesh for going through the draft and for providing valuable comments. We would like to thank the Chennai Mathematical Institute, Chennai, for the hospitality provided during the completion of this project.


\end{document}